\documentclass[aps,pre,twocolumn,superscriptaddress,showpacs]{revtex4-1}
\usepackage[english]{babel}
\usepackage[dvips]{graphics}
\usepackage{graphicx,epsfig}
\usepackage{amsmath}
\usepackage{color}
\usepackage{multirow}
\usepackage[normalem]{ulem}
\usepackage{booktabs}
\usepackage{lineno}
\usepackage{cancel}
\usepackage{hyperref}
\newcommand{\CR}[1]{{#1}}

\begin{document}

\title{The short periodic orbit method for excited chaotic eigenfunctions}
\author{F. Revuelta}
\email[E--mail address: ]{fabio.revuelta@upm.es}
\affiliation{Grupo de Sistemas Complejos, 
  Escuela T\'ecnica Superior de Ingenier\'\i a 
  Agron\'omica, Alimentaria y de Biosistemas,
  Universidad Polit\'ecnica de Madrid,
  Avda.~Puerta de Hierro 2-4,
  28040 Madrid, Spain.} 
%
\author{E. Vergini}
\email[E--mail address: ]{vergini@tandar.cnea.gov.ar}
\affiliation{
Departamento de F\'isica, Comisi\'on Nacional de Energ\'ia At\'omica,
Av.~del Libertador 8250, 1429 Buenos Aires, Argentina}
\author{R. M. Benito}
\email[E--mail address: ]{rosamaria.benito@upm.es}
\affiliation{Grupo de Sistemas Complejos, 
  Escuela T\'ecnica Superior de Ingenier\'\i a 
  Agron\'omica, Alimentaria y de Biosistemas,
  Universidad Polit\'ecnica de Madrid,
  Avda.~Puerta de Hierro 2-4,
  28040 Madrid, Spain.}
\author{F. Borondo}
\email[E--mail address: ]{f.borondo@uam.es}
\affiliation{Instituto de Ciencias Matem\'aticas (ICMAT), 
   Cantoblanco, 28049  Madrid, Spain.} 
\affiliation{Departamento de Qu\'imica, 
  Universidad Aut\'onoma de Madrid, 
  Cantoblanco, 28049 Madrid, Spain.}
\date{\today}

\begin{abstract}
In this paper [published in \
\href{https://journals.aps.org/pre/abstract/10.1103/PhysRevE.102.042210}{\emph{Phys. Rev. E} \textbf{102}, 042210 (2020)}],
a
new method for the calculation of excited chaotic eigenfunctions in arbitrary energy windows
is presented.
We demonstrate the feasibility of using wavefunctions localized on unstable 
periodic orbits as efficient basis sets for 
this task in
classically chaotic systems.
The number of required localized wavefunctions is only of the order of the ratio
$t_H/t_E$, with 
$t_H$ the Heisenberg time and $t_E$ the Ehrenfest time.
As an illustration, we present convincing results for a coupled two-dimensional quartic oscillator with 
chaotic dynamics. 
\end{abstract}

\maketitle

\section{Introduction} \label{sec.intro}
%
The accurate computation of the eigenstates of a dynamical system 
is a central problem in Physics and Computational Chemistry.
Usual standard methods \cite{Shao} 
are based on the variational principle, 
which implies that in order to get an approximation to the 
$N^\textrm{th}$ eigenstate, the $N-1$ lower lying ones should also be calculated.
This makes the task particularly demanding when interested
in excited states.
The existence of 
\CR{a large density of states} or classically chaotic dynamics
also contribute to significantly increment the computational burden.
In the case of systems with hard wall boundaries, such as quantum billiards,  
eigenstate computations are not performed by the variational procedure, 
and then efficient methods to calculate only the states within specific energy 
windows have been developed \cite{scaling}.
This fact dramatically reduces the computational time for arbitrarily excited levels, 
and that fully opened the possibility of studying these interesting states.
Unfortunately, such effective methods are not readily available for systems with 
continuous potentials.

In this paper, we present a method which extends the procedure recently 
developed by us~\cite{Revuelta13,Revuelta17} for the computation of 
eigenstates of classically chaotic systems. 
This procedure is based on the use of wavefunctions localized on unstable periodic 
orbits (POs), which leads to small basis sizes, of the order of the ratio  $t_H/t_E$
\CR{between the Heisenberg,~$t_H$, and the Ehrenfest,~$t_E$, times.
Recall that while the former provides the time scale at which the eigenenergies 
are accurately obtained,
which scales as~$t_H \sim {\mathcal O}(\hbar^{1-d} )$,
being $d$ the number of 
degrees of freedom of the system,
the latter gives the time lapse where the semiclassical 
evolution of a wave packet is valid,
being~ $t_E\sim{\mathcal O}(\log \hbar)$.
A specific definition of these two important characteristic times for our system 
will be given below in Sec.~\ref{sec:system}.}

The efficiency of the method is demonstrated by systematic application to 
small energy windows defined in different energy ranges of a
two-dimensional coupled quartic oscillator with highly chaotic dynamics.
The key point for the success of our method is the strong energy
localization of our basis functions.
\CR{This interesting property has enabled the calculation of the eigenfunctions 
of other chaotic systems, such as billiards~\cite{scaling}, 
maps~\cite{Ermann08}, or molecules~\cite{Revuelta16, Revuelta17},
which makes us to expect the general validity of our approach.}
Other attempts to use \CR{phase space} 
basis functions
have been made previously in the literature,
especially in relation with molecular systems.
\CR{For example, Davis and Heller used nonorthogonal complex Gaussians,
which render analytical expressions for the elements of the Hamiltonian matrix 
in the case of polynomial potentials~\cite{DavisHeller}.
Reimers and Heller extended that semiclassical basis sets for rigid 
rotors~\cite{ReimersHeller}.
Wavelets defined in the phase space have been also successfully applied in a series 
of papers by Poirier and coworkers~\cite{Poirier04a, Poirier04b, Halverson14}.
Similarly, phase space Gaussians were also applied, on the one hand, by 
Halverson and Poirier~\cite{Halverson12},
by incorporating some of their ideas previously used with wavelets,
and, on the one hand,  
by Shimshovitz, B\v{a}ci\'c and Tannor~\cite{Shimshovitz12, Shimshovitz14},
using the von Neumann lattice.
More recently, Brown and Carrington~\cite{BrownCarrington15, BrownCarrington16}
have developed an iterative eigensolver which boosts the numerical calculations.
Other long standing methods for computing molecular energies are the Discrete Variable 
Representation (DVR)~\cite{DVR1,DVR2} using localized basis elements,
for which the kinetic and potential energy contributions are easily computed,
and the use of the powerful Lanczos diagonalization method \cite{Lanczos}.}

Unstable POs are a cornerstone in classical and quantum 
chaotic dynamics~\cite{Gutzwiller90}.
They are the only remnants of order in the chaotic sea,
and their invariant manifolds organize the classical dynamics
in their neighborhoods~\cite{LL10}.
From the quantum mechanical point of view, 
POs have also been shown to have a deep impact on the energy 
levels of classically chaotic systems~\cite{Gutzwiller90}, 
and also on the characterization of many of their eigenfunctions,
where they can yield ``scars'' on the probability distributions,
as shown by Heller in his seminal paper~\cite{Heller84}.
Bogomolny showed that this localization phenomenon can
also arise in the combination of groups of eigenfunctions~\cite{Bogomolny88}, 
and later Berry developed the 
corresponding phase space version of this theory~\cite{Berry89}.

Different methods have been reported in the literature for 
the construction of localized states over unstable POs
(hereafter called ``scar functions'' due to its similarity
with Heller's scarring on eigenfunctions).
Some use averages over groups of eigenfunctions around the PO
quantization condition~\cite{Polavieja94};
others make use of short PO theory~\cite{Vergini01, Vergini02},
apply the asymptotic boundary layer~\cite{Vagov09},
or perform the quantum propagation of wavepackets launched
along the PO of interest~\cite{Revuelta12, Revuelta16}.

In this work, we discuss the spectral properties of our basis functions,
and demonstrate the accuracy of the proposed method by carrying out a systematic study
of the convergence, both for eigenenergies and eigenfunctions in a small energy window.
We also show the feasibility of using remarkably
small basis sizes to calculate arbitrarily excited states.  
{Only a brief description of the most relevant computational 
details and results are given in the text.}
Let us remark that the method reported would render even more accurate results 
in the case of more generic systems that present regular and irregular motion at the 
same time because of the more simple nodal structure of their eigenstates.

The rest of this work is organized as follows.
First, we briefly describe in the next section the system under study,
which is an anharmonic potential.
Second, we describe in Sec.~\ref{sec:method} the method
that we have develop to construct efficient basis sets,
in the sense that their sizes are small compared to the number
of accurately computed eigenfunctions.
Third, Sec.~\ref{sec:results} is devoted to the results and the corresponding discussion.
Finally, we sum up the paper in Sec.~\ref{sec:conclusions} with the conclusions and outlook.

\section{System} \label{sec:system}
%
The system we have chosen to study is 
described by the classical Hamiltonian
%
\begin{equation} \label{eq:H}
  \mathcal{H} (p_x, p_y, x, y)  = \frac{  p_x^2 + p_y^2 }{2} 
   + \frac{ x^2 y^2}{2} + \frac{x^4+ y^4}{400}.
\end{equation}
The dynamics of this system is extremely chaotic,
and no signs of regularity in phase space are observed 
at first sight~\cite{LL10}.
\par
The eigenfunctions of the associated quantum Hamiltonian belong to the five 
irreducible representations of the~$C_{2v}$ symmetry group. 
Nevertheless, we only consider here the totally symmetric
(even on the axes,~$x$ and~$y$, and the diagonals,~$x\!=\!\pm y$)
eigenfunctions, which belong to the $A_1$ representation.
\CR{The} mean number of $A_1$ eigenenergies smaller than $E$ 
is semiclassically estimated \CR{using the method reported in the Ref.~\cite{Bohigas93}
as
$\mathcal{N}(E) \simeq 0.251 E^{3/2} /\hbar^{2} +0.605 E^{3/4} /\hbar$.
Then, the mean energy density given by 
$\rho(E) = {\rm d}\mathcal{N}(E)/{\rm d}E \simeq 0.376 E^{1/2} /\hbar^{2} + 0.454 E^{-1/4}/\hbar $, while}
%
the Heisenberg time is defined by the relationship
\begin{equation}
  \CR{t_H=2 \pi \hbar \rho(E).}
\end{equation}

\begin{figure}
\includegraphics[width=0.8\columnwidth]{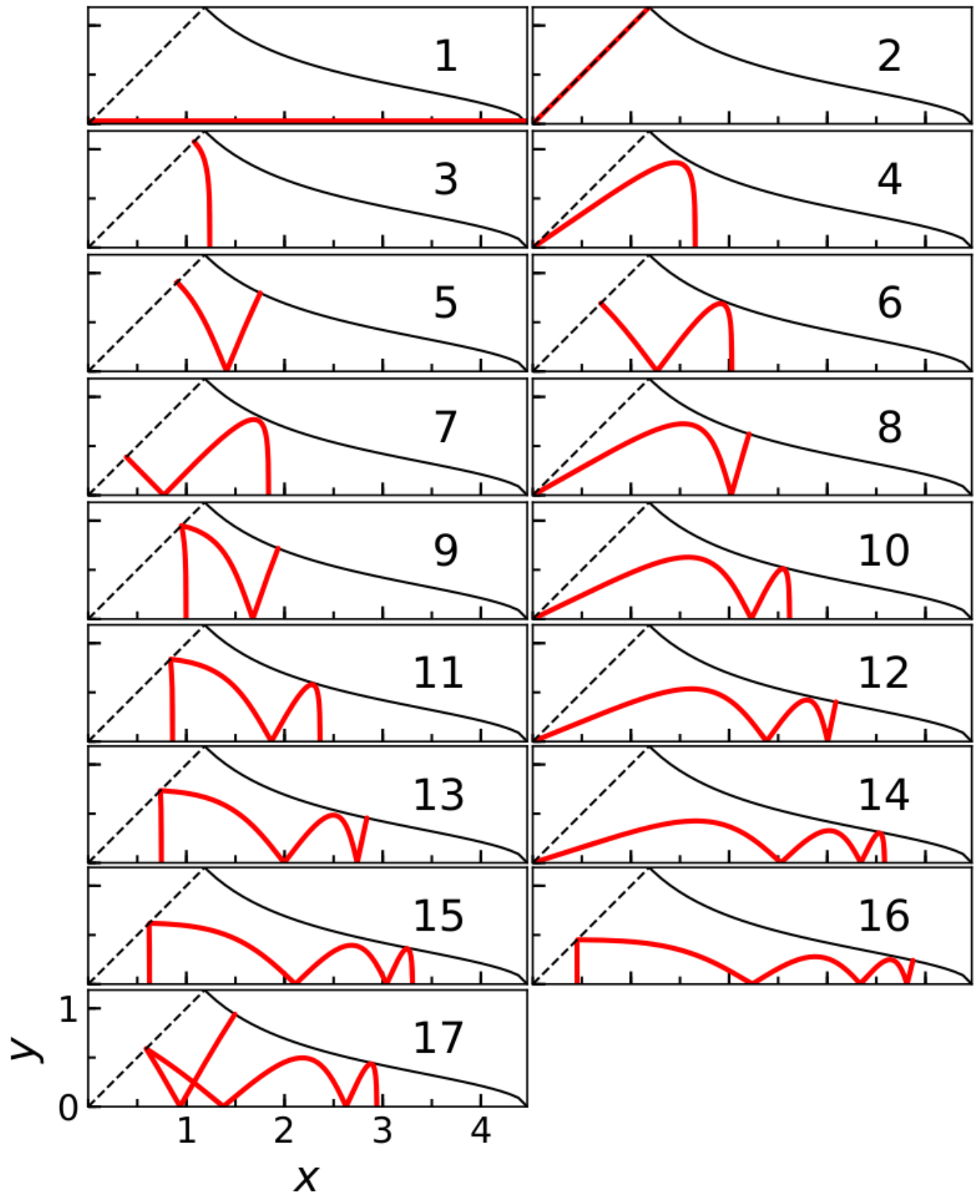}
  \caption{Set of desymmetrized periodic orbits used in the calculation (red curves). 
  The equipotential at energy $E=1$
  for Eq.~\eqref{eq:H}
  is shown in solid black curve.}
  \label{fig:1}
\end{figure}

For the construction of the semiclassical basis one selects a set of unstable short POs  in such a 
way that the sum of their periods
is greater than $t_H$ \cite{Ver0}
\begin{equation}
   \sum_\gamma T_{\gamma}  >  t_H,
\label{SPO}
\end{equation}
being $T_{\gamma}$ the period of a short PO $\gamma$.
\CR{At this point it is worth noticing that the monodromy method~\cite{Baranger88, Davies92} 
is exceptionally fast
for calculating periodic orbits in a general Hamiltonian system, specially in
high dimension. Nevertheless, here the situation is much more simple because
our Hamiltonian has only two degrees of freedom and it presents the
symmetries of the~$C_{4v}$ point group.
As a consequence, the shortest periodic orbits,
which are  symmetrical, are trivially evaluated. 
Furthermore, the
Hamiltonian~\eqref{eq:H} is mechanically similar, and then it presents
no bifurcations as a function of the energy. 
Thus, the phase
space structure does not change with the energy, and then one can compute all
the periodic orbits at a reference energy,~$E$, and scale them for
a different energy,~$E'$, as 
\begin{eqnarray} x'_{t'} & = & \left( \frac{E'}{E} \right)^{1/4} x_t, \quad 
  p'_{x', t'} = \left( \frac{E'}{E} \right)^{1/2}
  p_{x, t}, \label{eq:xpx} \nonumber\\ 
  y'_{t'} & = & \left( \frac{E'}{E} \right)^{1/4} y_t, \quad 
  p'_{y', t'} = \left( \frac{E'}{E} \right)^{1/2} p_{y, t}, 
  \label{eq:ypy} 
\end{eqnarray} 
with~$t' = \left( E' / E \right)^{-1/4} t$. 
Of course the first two periodic orbits of Fig.~\ref{fig:1} are given without
any calculation and consequently let us consider, for instance, orbits~3,~6
and~7. By fixing the starting point with $y=0$ and $p_x=0$ one looks for the
coordinate $x$ in such a way that the first intersection of the trajectory
with the diagonal occurs orthogonal to it.}

In this paper, we use the POs displayed in Fig.~\ref{fig:1},
with $\sum T_{\gamma} \simeq 32$  and $t_H\simeq 26$ at $E=110$, 
the maximum energy considered in the calculation. 
These POs are arranged by the number of bounces with the symmetry lines.
Obviously, a criterion of general validity for arranging short POs with the purpose 
of covering phase space in an efficient way is desirable. 
We believe that such criterion could be defined in terms of the properties
of the first homoclinic orbit of each short PO,
particularly in higher dimension.
Still, the condition~\eqref{SPO}
and the fact that one only needs the shortest POs of the system
strongly limits the time required to search for POs.
Moreover, symmetry considerations can further simplify this task.

Associated with an unstable short PO $\gamma$, one obtains the discrete set 
of Bohr-Sommerfeld (BS) energies, $E^{\gamma}_n$, computed from the relationship
\begin{equation}
  \frac{S_{\gamma}}{\hbar} - \frac{\mu_{\gamma} \pi}{2} = 2 \pi n,
\end{equation}
 with $ S_{\gamma}$ the action of $\gamma$ and $\mu_{\gamma}$ its Maslov index. 
\par
We provide a semiclassical construction of wavefunctions with $n$ excitations along $\gamma$, 
the so-called tube functions of $\gamma$, with mean energy $E^{\gamma}_n$
and small energy dispersion given by $ \hbar \lambda_{\gamma}/\sqrt{2}$; where $\lambda_{\gamma}$ is
the stability exponent of $\gamma$. 
Although the time evolution of these wavefunctions can be evaluated 
at semiclassical level in terms of homoclinic and heteroclinic orbits \cite{Ver1, Ver2}, 
here we prefer to use a quantum evolution, following Ref. \cite{Sibert}, 
in order to obtain highly accurate eigenfunctions and eigenvalues.
In this respect, it is worth mentioning that the quantum evolution of tube functions up to times
of the order of the mean Ehrenfest time $t_E$, that in our case is given by
\begin{equation}
  t_E=\frac{1}{2 \lambda} \ln \left(\frac{\mathcal{A}}{\hbar}\right),
\end{equation}
with $\lambda\simeq 0.385 E^{1/4}$ the Lyapunov exponent of the system, and
$\mathcal{A} \simeq 11.1 E^{3/4}$ the area of a characteristic Poincar\'e 
surface of section,  is a simple task.

\section{Method} \label{sec:method}
%
For the computation of eigenvalues and eigenfunctions 
contained in an energy window
defined by the central energy $E_0$ and width $\Delta E$, $[E_0-\Delta E  ,E_0+ \Delta E ]$, 
we proceed as follows. 
\par
First, we select the set of BS energies of the POs
existing in the \textit{enlarged} energy
window~$[E_0-(\Delta E+ \delta), E_0+(\Delta E +\delta)]$, with
$\delta \simeq 2 (t_H/t_E)/ \rho(E)$ an energy window containing
$2 t_H/t_E$ eigenvalues.
\par
Second, we construct tube functions of the selected BS energies, for instance $E^{\gamma}_n$.
By propagating a Gaussian wave packet over $\gamma$ 
using the frozen Gaussian approximation~\cite{Heller75}, one has
\begin{equation} \label{eq:tube}
  \psi_n^{\gamma} (x,y) = \int_0^{T_{\gamma}} 
    e^{i E_n^{\gamma} t / \hbar} \phi(x, y, t)  {\rm d}t,
\end{equation}
where
\begin{eqnarray}
\!\! \! \phi(x,y,t) & \!=& \exp\left\{-\alpha  [(x-x_t)^2+ (y-y_t)^2 ] + \right . 
           \nonumber \\
  &   & \!\!\!\!\!\! \left. i [{p_x}_t (x-x_t)/\hbar+{p_y}_t (y-y_t )/\hbar+ 2 \pi n t /T_{\gamma}] \right\}, \nonumber \\
  \label{eq:CS}
\end{eqnarray}
with  $(x_t, y_t, {p_x}_t, {p_y}_t)$ the classical evolution over $\gamma$.
\CR{The energy dispersion of $\psi_n^{\gamma}(x,y)$ behaves like
$C \hbar$ in leading order, with $C$ a constant that depends
on the parameter~$\alpha$ of Eq.~(\ref{eq:CS}). By selecting the
parameter $\alpha$ that minimizes the  energy dispersion for
a finite value of $\hbar$, we obtain a good estimate of the
parameter that minimizes the coefficient $C$.
Of course, the main idea behind this choice is that as the energy
dispersion is smaller, the wave function is closer to an
eigenfunction.}
In conclusion, one selects the parameter
$\alpha$ in such a way that the energy dispersion of $ \psi_n^{\gamma} (x,y) $ 
results
a minimum.
\par
\begin{figure}
 \includegraphics[width=.80\columnwidth]{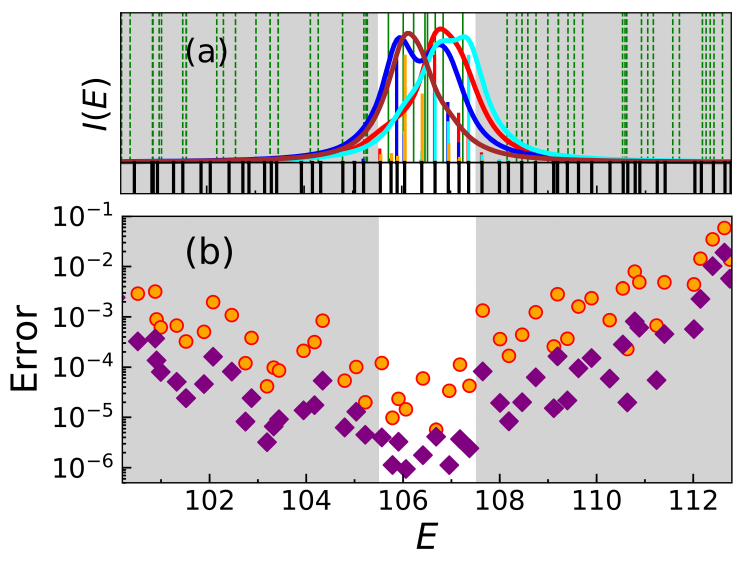}
  \caption{
\CR{(a) Bohr-Sommerfeld quantized energies 
(vertical green lines) for the 61 scar functions in the basis set used
to obtain the 9 eigenstates of $A_1$ symmetry
(vertical black lines) of the quartic oscillator~\eqref{eq:H}
lying in the energy window [105.5, 107.5]
(between gray shaded regions).
The infinite (vertical colored lines)
and low resolution
(continuous thick curves)
 projections on the eigenenergies spectrum
of the
$\vert 11, 44 \rangle$ (red),
~$\vert 5, 24 \rangle$ (blue),
~$\vert 14, 55 \rangle$ (cyan),
~$\vert 17, 67 \rangle$ (brown)
 scar functions shown in Fig.~\ref{fig:3}
(see text for notation),
which are the ones which are mainly involved 
in the reconstruction of the eigenfunction~297,
have been superimposed.
(b) Errors in the computed eigenenergies 
(orange circles) and eigenfunctions (purple diamonds) given 
by Eq.~\eqref{eq:error}.}
}
 \label{fig:errors}
\end{figure}

Third, {\it scar functions} are constructed by quantum  propagation 
of the tube functions~\eqref{eq:tube} up to $t_E$, 
%
\begin{equation} \label{eq:scar}
  \langle x,y | \gamma,n \rangle \! =\!  \int_{-t_E}^{t_E} \!  
    \! \!    \cos \left( \frac{\pi t}{2 t_E} \right) \;
    \! \! e^{i
    \left(E_n^{\gamma} -\mathcal{\hat{H}}\right)
     t /\hbar} 
     \psi_n^{\gamma}(x,y) {\rm d}t .
\end{equation}
This time evolution is the most demanding point of our method,
but still it can be efficiently carried out using procedures like the 
discrete variable representation~\cite{DVR1,DVR2} because
of the small energy dispersion of the tube functions \cite{Sibert}.
Similar results are expected when 
other basis functions with similar spectral properties are used, e.g.
the scar functions constructed using the method reported in Ref.~\cite{Vergini01}.
\CR{Similarly, Eq.~\eqref{eq:tube} could be also used
to construct the semiclassical basis set, but the (much) smaller spectral dispersion
of the scar functions~\eqref{eq:scar} after a relative short time propagation
makes these last functions better suited.
As a particular example, notice  in Fig.~\ref{fig:errors}(a) that the
spectra of four of the scar functions involved in our calculations is mostly concentrated 
in the window of interest.
Here, the infinite resolution spectra are shown as vertical thin lines,
while its finite-resolution counterpart, which is defined as the Lorenzian convolution
\begin{equation}
 I_\Delta ( E ) =  \frac{\Delta E}{2\pi } \sum_{N}  \frac{\langle \psi_n^\gamma \vert N \rangle^2}
                                                       {\left( E - E_{N} \right)^2 + \left(\Delta E/2\right)^2},
\label{eq:I}
\end{equation}
with~$\Delta E = 0.4$ and~$E_{N}$ is the energy
of the eigenstate~$\vert N \rangle$,
are shown as vertical black thick lines at the bottom of Fig.~\ref{fig:errors}(a).
Notice that this spectra can only be obtained once the
eigenfunctions of the system have been calculated.
}
%

\par
Fourth, the Gram-Schmidt Selective method~\cite{Revuelta13} is then used 
to choose the most fit scar functions.
The method works as follows. One selects the scar function with
minimum energy dispersion as the first function of the basis, for instance $| \gamma,n \rangle $.  
Next, one eliminates, for the other scar functions, the contribution along $| \gamma,n \rangle $, for instance  
$| \eta ,m \rangle $ (with $\eta$ other PO) is replaced by $| \eta ,m \rangle  - \langle \gamma, n | \eta , m \rangle | \gamma, n \rangle $.
Then, one evaluates the energy dispersion of the new wave functions and selects as the second 
function of the basis, the one with
minimum dispersion. This process continues in a similar way  to the Gram-Schmidt method
while the minimum energy dispersion 
is smaller than a given value, for instance~$2 \delta$. The method 
has the effect of
describing  the subspace 
spanned by the desired 
eigenfunctions, i.e.,~those
whose energies lie within the energy window under study.

Fifth, once the final basis set is obtained
the corresponding Hamiltonian matrix is computed and 
diagonalized using standard procedures.
\begin{figure}
 \includegraphics[width=.80\columnwidth]{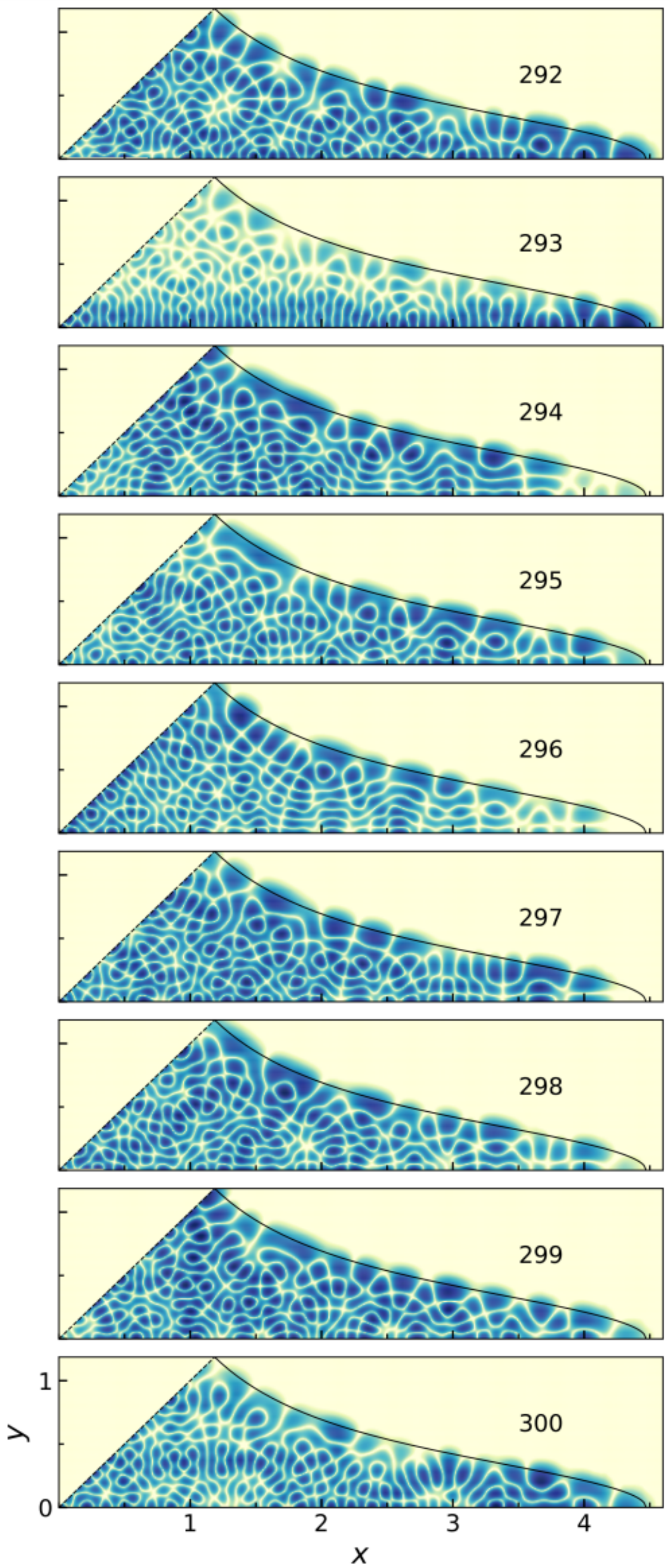}
  \caption{Probability density of the eigenfunctions contained  
   in the energy window $E=[105.5,107.5]$. 
  They are scaled to $E=1$ for comparison.
  The corresponding eigenstate number is given in the plots.}
 \label{fig:2}
\end{figure}

\section{Results and discussion} \label{sec:results}
%
Figure~\ref{fig:2} presents the results 
of a computation performed in the energy window 
$[105.5,107.5]$
\CR{for~$\hbar=1$}.
There are $66$ scar functions contained in the enlarged energy window
$[100.2, 112.8]$ \CR{[see vertical green lines in Fig.~\ref{fig:errors}(a)]},
and the Gram-Schmidt Selective method chooses $61$ scar functions
in order
to reproduce with high accuracy the $9$ eigenvalues
\CR{[black lines contained in the central white region of Fig.~\ref{fig:errors}(a)]}
 and eigenfunctions
of interest.
The figure displays the probability density of
these eigenfunctions in the fundamental domain.
As can be seen, they all present a very complex 
nodal pattern, something characteristic of classically chaotic systems. With
respect to the scar phenomenon, we clearly observe that the eigenfunction $293$ 
is localized along the first PO of Fig.~\ref{fig:1}.
\par
The eigenfunction $|j\rangle$ of our computation is 
compared  with the ``exact'' result $|j'\rangle$ 
obtained using a basis 
set of $5000$ harmonic oscillator functions. We compute the error of eigenvalues
and eigenfunctions  by the dimensionless quantities
\begin{equation} \label{eq:error}
  \rho(E)\, |E_j-E_{j'}| \quad {\rm and} \quad 
   1 - \langle j \vert j' \rangle^2,
\end{equation}
respectively. 
The mean error of the $9$ eigenvalues and eigenfunctions of Fig.~\ref{fig:2}
is 
of order~$\sim10^{-4}$ and $\sim10^{-5}$,
respectively, 
\CR{
as shown in Fig.~\ref{fig:errors}(b).
A similar accuracy 
based on the variational principle
would require much larger full matrix diagonalizations.
In particular, for the energy window reported in the manuscript, the same
level of accuracy
%
requires a harmonic oscillator
basis set formed by $\sim50$ times more basis
elements than the one 
reported here,
which exerts a diagonalization
time $\sim50^3$ times larger.}
Furthermore, by accepting an
error ten times greater than the previous ones, the same calculation provides $20$ consecutive 
eigenfunctions. With respect to the stability of the method, 
let us remark that in other energy windows 
of the spectrum up to $E=110$, we found essentially the the same result.
Moreover, for smaller energies a reduced number of POs is
required as long as Eq.~(\ref{SPO}) is accomplished.
Finally, we find that the dimension of the basis is only 40\% greater 
than the number of eigenfunctions contained in the enlarged energy window. 

Let us consider now the  representation of the
eigenfunctions obtained in our calculation in the basis set of scar functions,
this showing the basis elements that contribute the most (largest overlaps)
to their reconstruction.
Notice that this procedure provides valuable information
on the influence that each PO has on the emergence of the 
properties of individual chaotic eigenfunctions,
this representing an important added bonus of our method 
to the theory of \textit{quantum chaos}.
%
\begin{figure}
  \includegraphics[width=.80\columnwidth]{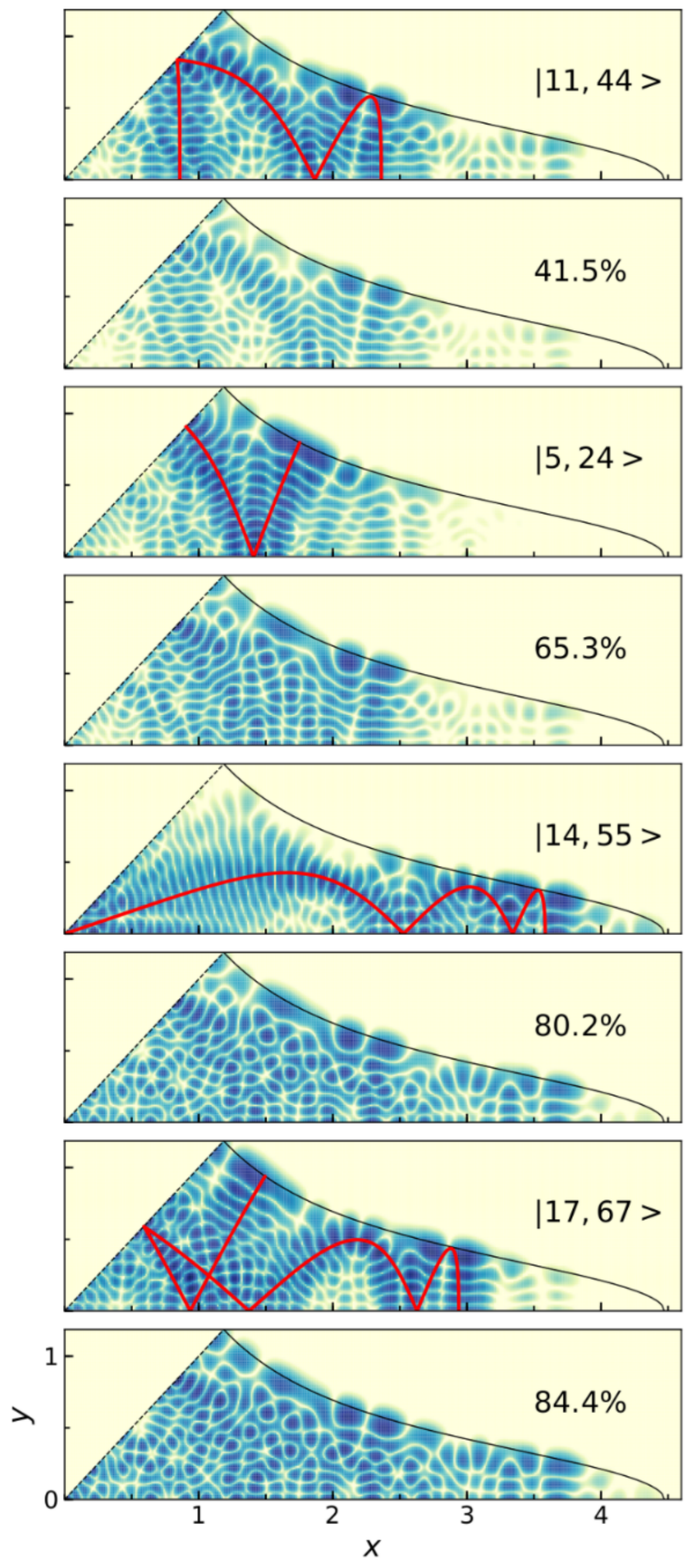}
  \caption{Partial reconstruction of eigenfunction~$297$ (see Fig.~\ref{fig:2})
   by incorporating the most relevant scar functions.
   The percentage of reconstruction at each step
   is also given.
}
\label{fig:3}
\end{figure}
As an illustrative example, we present in Fig.~\ref{fig:3} the 
reconstruction of the eigenfunction~$297$ shown in 
Fig.~\ref{fig:2}.
As already mentioned, it exhibits a complex nodal pattern,  
as it is not localized over any single PO.
Successive steps in the reconstruction are given in the figure.
In each one of them we show the result of adding the 
scar function depicted in the panel above it.
Notice that 41.5\% of the eigenfunction is reconstructed solely with the scar 
function~$\vert 11, 44 \rangle$.
By adding the scar function~$\vert 5, 24 \rangle$,
65.3\% of the eigenfunction is reconstructed.
This figure increasingly raises when more basis functions
are included, up to a value of 84.4\% when four basis functions are combined. 
In other words, the square modulus of the projection of the eigenstate 297 
onto the subspace spanned by the four scar functions of the figure is  $0.844$.
Notice also how the reconstruction procedure, that we have
just described, progressively refines the intricate details
of the eigenfunction.

The other eigenfunctions of Fig.~\ref{fig:2} have a similar partial reconstruction
in terms of a small number of scar functions. In particular, we remark the 
cases where only one scar function is able to reproduce more than 50\% of
the eigenfunction. The scar function~$\vert 1, 56 \rangle$ reproduces 79.7\% of
293,~$\vert 8, 41 \rangle$ reproduces 60.6\% of 296, and~$\vert 16, 56 \rangle$
reproduces 58.9\% of 300.

\begin{figure}
  \includegraphics[width=.8\columnwidth]{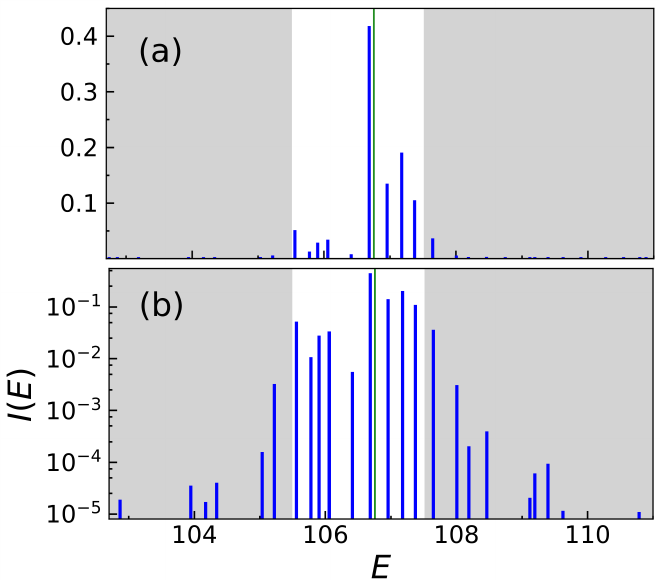}
  \caption{
\CR{Square modulus of the overlap of the
first wave function of Fig.~\ref{fig:3}
, i.\,e., the scar function $\vert 11, 44 \rangle$,
with the eigenfunctions of
the system, as a function of the eigenenergies
in linear~(a)  and logarithmic~(b) scales.
The thin green line marks the position of the Bohr-Sommerfeld
quantization energy ($E\simeq 106.75$),
which lies within
the reference energy window [105.5, 107.5] 
(between the gray shaded regions).}
}
  \label{fig:5}
\end{figure}

\CR{In order to clearly show the strong energy localization of our basis functions, 
we conclude this section by displaying
in Fig.~\ref{fig:5} the square modulus of the overlap of the 
first wave function in Fig.~\ref{fig:3}, 
i.\,e., the scar function $\vert 11, 44 \rangle$,
with the eigenfunctions of the system.
In panel (a) we use a linear scale to remark the fact that the energy dispersion of 
the wave function consists only of contributions within a few mean level spacings. 
Notice here that the highest peak 
 lies 
very close to the Bohr-Sommerfeld energy of
the scar function (thin green line at~$E\simeq106.75$), and
corresponds to the
overlap associated with the eigenstate~297.
Recall that the scar function $\vert 11, 44 \rangle$
is precisely  
the main contribution to the reconstruction of that eigenstate.
Furthermore, in panel~(b) we use a logarithmic scale to emphasize that the
intensities decay exponentially as we move away from the corresponding
Bohr-Sommerfeld energy. 
}
%

\section{Conclusions and Outlook} \label{sec:conclusions}
The short periodic orbit method, formulated and tested in 
this 
paper, 
is an extremely powerfull technique for calculating 
very excited chaotic eigenfunctions. 
We have shown the 
feasibility of computing eigenfunctions of a classically
chaotic system using small basis sets.
The key point 
is the use of scar functions
constructed in the neighborhood of a set of short periodic orbits that covers 
the relevant phase space efficiently.
These semiclassical functions have excellent properties 
regarding localization in energy that make them suitable to describe 
only the portion of the system's Hilbert space
spanned by the eigenstates of interest.
In particular, the accuracy of the method has been demonstrated by
performing a systematic calculation of 
the eigenfunctions of a coupled quartic oscillator
using narrow energy windows in excited regions of the spectrum.
Let us finally remark that each eigenfunction
obtained in the calculation is partially reconstructed using
only a  few scar functions,
of the order of the~$t_H/t_E$ ratio.
This fact provides a useful way of classifying chaotic eigenfunctions.

\begin{acknowledgments}
This work has been partially supported by the 
Ministerio de Econom\'ia, Competitividad e Innovaci\'on
(MINECO)
of the Gobierno de Espa\~na
under Contract No. MTM2015-63914-P, 
by the Ministerio de Ciencia, Innovaci\'on y Universidades
under Contract No. PGC2018-093854-B-I00
by ICMAT Severo Ochoa under Contract SEV-2015-0554,
F.R. also acknowledges the financial support of the
programs to improve the research of young doctors
of the Programa Propio of the Universidad Polit\'ecnica de
Madrid,
and of the project GeoCoSiM, financed under the
Plurianual Agreement between the Comunidad de Madrid and  
the Universidad Polit\'ecnica de Madrid.

\end{acknowledgments}



\end{document}